\documentclass[twocolumn,aps,prl,superscriptaddress,showpacs,floatfix]{revtex4}
\newcommand{\nc}{\newcommand}
\nc{\la}{\left\langle}
\nc{\ra}{\right\rangle}

\usepackage{amsmath,bm}
\usepackage{graphicx}

\begin{document}

\title{Scalings of Elliptic Flow for a Fluid at Finite Shear Viscosity}
\author{G. Ferini}
\affiliation{INFN-LNS, Via S. Sofia 62, I-95125 Catania, Italy}
\author{M. Colonna}
\affiliation{INFN-LNS, Via S. Sofia 62, I-95125 Catania, Italy}
\author{M. Di Toro}
\affiliation{INFN-LNS, Via S. Sofia 62, I-95125 Catania, Italy}
\affiliation{Dipartimento di Fisica e Astronomia, Via S. Sofia 64, I-95125 Catania, Italy}
\author{V. Greco}
\affiliation{INFN-LNS, Via S. Sofia 62, I-95125 Catania, Italy}
\affiliation{Dipartimento Interateneo di Fisica di Bari, Via Amendola 173, I-70126 Bari, Italy}

\date{\today}

\begin{abstract}
Within a parton cascade approach we investigate the scaling of the differential elliptic flow $v_2(p_T)$ with eccentricity $\epsilon_x$ and system size 
and its sensitivity to finite shear viscosity. We present calculations for shear viscosity to entropy density ratio $\eta/s$ in the range from $1/4\pi$ up to $1/\pi$, finding that the 
$v_2$ saturation value varies by about a factor 2. 
Scaling of $v_2(p_T)/\epsilon_x$ is seen also for finite $\eta/s$ which indicates that it does not prove
a perfect hydrodynamical behavior, but is compatible with a plasma at finite $\eta/s$.
Introducing a suitable freeze-out condition, we see a significant reduction of 
$v_2(p_T)$ especially at intermediate $p_T$ and for more peripheral collisions.
This causes a breaking of the scaling for both $v_2(p_T)$ and the $p_T-$averaged $v_2$,
while keeping the scaling of $v_2(p_T)/\la v_2\ra$. This is in better agreement with the experimental
observations and shows as a first indication that the $\eta/s$ should be significantly lower than the pQCD
estimates.
We finally point out the necessity to include the hadronization via coalescence for a definite evaluation
of $\eta/s$ from intermediate $p_T$ data. 

\end{abstract}

\maketitle
The Relativistic Heavy Ion Collider (RHIC) has successfully shown that a transient state of matter
at initial temperature T and energy density $\epsilon$ well above the one expected at the phase
transition (T$_c$ $\sim$ 170 MeV and $\epsilon_c \sim 0.7$ GeV/fm$^3$) has been created.
In particular the large value of the elliptic flow $v_2$ indicates that such a matter, called quark-gluon
plasma (QGP), behaves like a nearly perfect fluid. In fact the dynamics of the bulk of plasma (i.e. for transverse momentum $p_T < 1.5$ GeV)  is successfully described by ideal hydrodynamics \cite{Kolb:2003dz},
at least for the most central collisions \cite{Ackermann:2000tr}.
At higher transverse momenta, due to incomplete equilibration,
the hydrodynamical behavior breaks down as confirmed by the saturation of the baryon to meson
ratio and by the quark number scaling of elliptic flow $v_2$ \cite{greco,greco2,fries,Molnar:2003ff,Hwa:2004ng}.
In this intermediate $p_T$ region ($1.5<p_T<5$ GeV) kinetic theory provides the most reliable approach and indeed parton cascade has successfully predicted the $v_2(p_T)$ saturation pattern
for $p_T \geq 1.5$ GeV \cite{moln02}. Furthermore the cascade approach hints at a parton cross 
section significantly larger than estimated in perturbative QCD (pQCD) in general consistency with 
the observed nearly hydrodynamical behavior.
On the other hand a minimum viscosity is imposed by quantum mechanical considerations \cite{Danielewicz:1984ww}
and more recently a study of supersymmetric gauge theory in infinite coupling limit \cite{Kovtun:2004de} has 
given a lower bound for the shear viscosity to entropy density ratio $\eta/s \geq 1/4\pi$. All known substances,
from water to a meson gas, obey this bound and 
indeed all of them are significantly above it \cite{Lacey:2006bc}. 
A first recent evaluation of shear viscosity in lattice QCD (lQCD) is consistent with the lower bound \cite{Nakamura:2004sy,Meyer:2007ic}  and show a 
mild evolution with temperature in the range of temperature covered by RHIC \cite{Meyer:2007ic}. 
The description of the elliptic flow, the strong scattering
of heavy quarks, the measurements of $p_T$ fluctuations, 
all point \cite{Lacey:2006bc} to a $\eta/s$ that should be close to the
bound and much smaller than the one expected in a pQCD regime \cite{Arnold:2003zc},
(about 5-10 times the lower bound).
A first attempt based on Knudsen number analysis of $v_2/\epsilon_x$ has lead to estimate 
$\eta_/s \sim 0.11-0.19$ depending on initial conditions \cite{Drescher:2007cd}.
Moreover it has been started an effort in developing viscous hydrodynamics \cite{Romatschke:2007mq,Song:2007fn} 
which is indicating a significant reduction of $v_2$ even for 
$\eta/s$ at the lower bound. The effect appears to be quite strong in the calculations of Ref. \cite{Song:2007fn} 
while is less pronounced in Ref. \cite{Romatschke:2007mq} where 
data on $p_T$ averaged elliptic flow, $\la v_2 \ra$, are fairly well reproduced  as a function of the 
collision centrality
with $\eta/s \sim 0.1$. On the other hand the low $p_T$ minimum bias $v_2(p_T)$ seems to favor calculations with an
even smaller $\eta/s$.
Similar findings for the dependence of the average elliptic flow $\la v_2 \ra$ on shear
viscosity have been reported also
in the context of a parton cascade \cite{Xu:2007jv}.
However a detailed investigation of viscous effects on differential elliptic flow $v_2(p_T)$ within a transport theory is still pending.

In this letter we present a study of the elliptic flow  and its scaling properties as a 
function of $p_T$ at finite $\eta/s$ in the range $(4\pi)^{-1}<\eta/s<\pi^{-1}$.
The analysis is based on a parton cascade approach and it is mainly focused on the intermediate $p_T$ region, 
where kinetic theory automatically accounts for non equilibrium effects. 
The main idea is to keep the $\eta/s$
of the medium constant during the collision dynamics. The parton cross section is rearranged according to the local density and momentum values.
Simulations have been carried out for a large range of impact parameters in 
$Au+Au$ collisions at $\sqrt{s_{NN}}= 200$ GeV. Some simulations have been performed also for $Cu+Cu$
for a first investigation of the system size dependence.
A first issue that we discuss is the scaling of the $v_2$ with the spatial eccentricity  
$\epsilon_x =\langle y^2 -x^2\rangle/\langle x^2+y^2\rangle$ and the system size.
We show that if $\eta/s$ is kept constant down to the thermal
freeze-out ($\epsilon \sim 0.2$ GeV/fm$^3$) a parton cascade exhibits a
$v_2/\epsilon_x$ scaling in the whole $p_T$ range investigated (up to 3.5 GeV). 
Therefore the prediction of the scaling is not a unique feature of ideal hydrodynamics
\cite{Bhalerao:2005mm}.
However experimentally the (in)dependence of $v_2/\epsilon_x$ 
on the centrality of the collision and on the system size is indeed a delicate issue
as raised by recent publications from PHENIX 
\cite{Adare:2006ti} and STAR \cite{:2008ed}. 
We point out that once a suitable freeze-out condition is introduced at $\epsilon_c \sim$ 0.7 GeV/fm$^3$
a cascade approach at finite viscosity can account for the breaking of the
scaling for $v_2(p_T)/\epsilon_x$ together with a persisting scaling for $v_2(p_T)/\la v_2\ra$, as 
experimentally observed. 

A first attempt of our investigation has been also to put a reasonable constraint on the $\eta/s$ value of the RHIC fluid through the observed
$v_2(p_T)$ pattern for 1 GeV$<p_T<$3 GeV. A significant dependence of $v_2(p_T)$ on shear viscosity is found with a reduction of the saturation value of nearly a factor 2 going from the lower bound to $\eta/s=\pi^{-1}$. 
However a definitive evaluation of $\eta/s$ is entangled with the observation of quark number scaling
in the same $p_T$ range and hadronization by coalescence plus fragmentation has to be self-consistently
included.

The partonic transport approach at the present stage does not contain the
different aspects of the dynamics and in particular it misses the effects of the fields, which have not yet been included
in a partonic transport code. However
it is certainly a powerful approach whenever a finite mean free path has to be considered and in particular at intermediate $p_T$ where the hydrodynamical behavior breaks down.
We have developed a $3+1$ dimensional Montecarlo cascade for on-shell partons
based on the stochastic interpretation of the
transition rate. Such an interpretation is free from several unphysical drawbacks and particularly suitable for an extension to multiparticle
collisions as pointed out by Z. Xu and C. Greiner \cite{Xu:2004mz}. The evolution of parton distribution function  from initial conditions through elastic scatterings is followed by propagating particles along straight lines and sampling  possible transitions in a certain volume and time interval according to the Boltzmann equation for two-body scatterings:
\begin{equation}
 p_{\mu} \partial^\mu f_1 \!=\! 
\int\limits_2\!\!\! \int\limits_{1^\prime}\!\!\! \int\limits_{2^\prime}\!\!
 (f_{1^\prime} f_{2^\prime}  -f_1 f_2) \vert{\cal M}_{1^\prime 2^\prime \rightarrow 12} \vert^2 
 \delta^4 (p_1+p_2-p_1^\prime-p_2^\prime)
\end{equation}
where $\int_j= \int_j d^3p_j/(2\pi)^3\, 2E_j $, $\cal M$ denotes the transition matrix for the elastic processes and $f_j$ are the particle distribution functions.

For the numerical implementation, we discretize the space into cells small respect to the system size and 
we use such cells to calculate all the local quantities. In particular we evaluate at each timestep the local collision probability and decide whether or not a collision can occur by means of a Monte Carlo algorithm. We
have performed several checks to test the validity of the code similarly to what thoroughly
discussed in Ref.\cite{Xu:2004mz}, obtaining the same results. More specifically we have performed tests
to choose a good discretization for convergency of the results for the elliptic
flow that is the main observable analyzed in the present paper.
The calculations shown are performed with cells of transverse area $ 0.5$ fm$^2$
and a longitudinal size of $\Delta \eta_s =0.1$, where $\eta_s$ is the space-time rapidity.
Furthermore we have implemented the subdivision (or test particles) technique which allows for a better mapping 
of the phase space. This is indeed necessary due to the smallness of the cell volume. Our tests have indicated that $N=6$ test particles are sufficient to achieve stable results for the collision rate and the elliptic flow.

In kinetic theory in ultra-relativistic conditions the shear viscosity can be expressed as \cite{degroot}
\begin{equation}
 \eta=\frac{4}{15} \rho \langle p \rangle \lambda
\label{eq1}
\end{equation}
with $\rho$ the parton density, $\lambda$ the mean free path and $\langle p \rangle$ the average momentum.
Therefore considering that the entropy density for a massless gas is $s=\rho(4-\mu/T)$, $\mu$ being
the fugacity, we get:
\begin{equation}
 \dfrac{\eta}{s}=\dfrac{4\langle p \rangle}{15\, \sigma_{tr} \rho (4 - \mu/T )}
\label{eq2}
\end{equation}
where $\sigma_{tr}$ is the transport cross section, defined as $$\sigma^{tr}=\int d\theta \frac{d\sigma}{d\theta} \texttt{sin}^2 \theta .$$ We use a pQCD inspired cross section with the infrared singularity regularized by 
Debye thermal mass $m_D$ \cite{moln02}:
\begin{equation}
 \frac{d\sigma}{dt} = \frac{9\pi \alpha_s^2}{\left(t+m_D^2\right) ^2}\left(\frac{1}{2}+\frac{m^2_D}{2 s}\right) 
\label{eq3}
\end{equation}
where $s,t$ are the Mandelstam variables and $m_D=$ 0.7 GeV.

Our approach is to artificially keep the viscosity of the medium constant during the dynamics of the collisions
in a way similar to \cite{Abreu:2007kv}.
This is achieved by evaluating
locally in space and time the strength of the cross section needed to keep the $\eta/s$ constant.
From Eqs. (\ref{eq1}) and (\ref{eq2}) we see that assuming locally the thermal equilibrium
this can be obtained evaluating in each cell the cross section according to:
\begin{equation}
 \sigma_{tr} = \dfrac{4 \,}{15} \dfrac{ \langle p \rangle}{ \rho (4- \mu/T)} \dfrac{1}{\eta/s},
\end{equation}
with $\eta/s$ set from 1 to 4 in units of the minimum value.

Partons are initially distributed according to a standard mixture of
the density of participant nucleons ($80\%$) and of binary collisions ($20 \%$) calculated with a standard Glauber model. 
The eccentricity of the system is therefore similar
to the one used in standard calculations. We also start our simulation like in hydrodynamics
at a time $t=0.6$ fm assuming for partons with $p_T < p_0=$2 GeV a thermalized spectrum and for $p_T >p_0$  the spectrum
of non-quenched minijets as calculated in \cite{Zhang:2001ce}.
In principle the transport approach allows for an investigation of the important issue of thermalization
which should be strictly related to three-body scatterings \cite{Xu:2004mz}. Here looking at 
collective modes like the elliptic flow we implicitly assume that the results do not depend
significantly on the details of the collision kinematics once the shear viscosity has been fixed.

\begin{figure}[h]
\includegraphics[height=1.7in,width=3.3in]{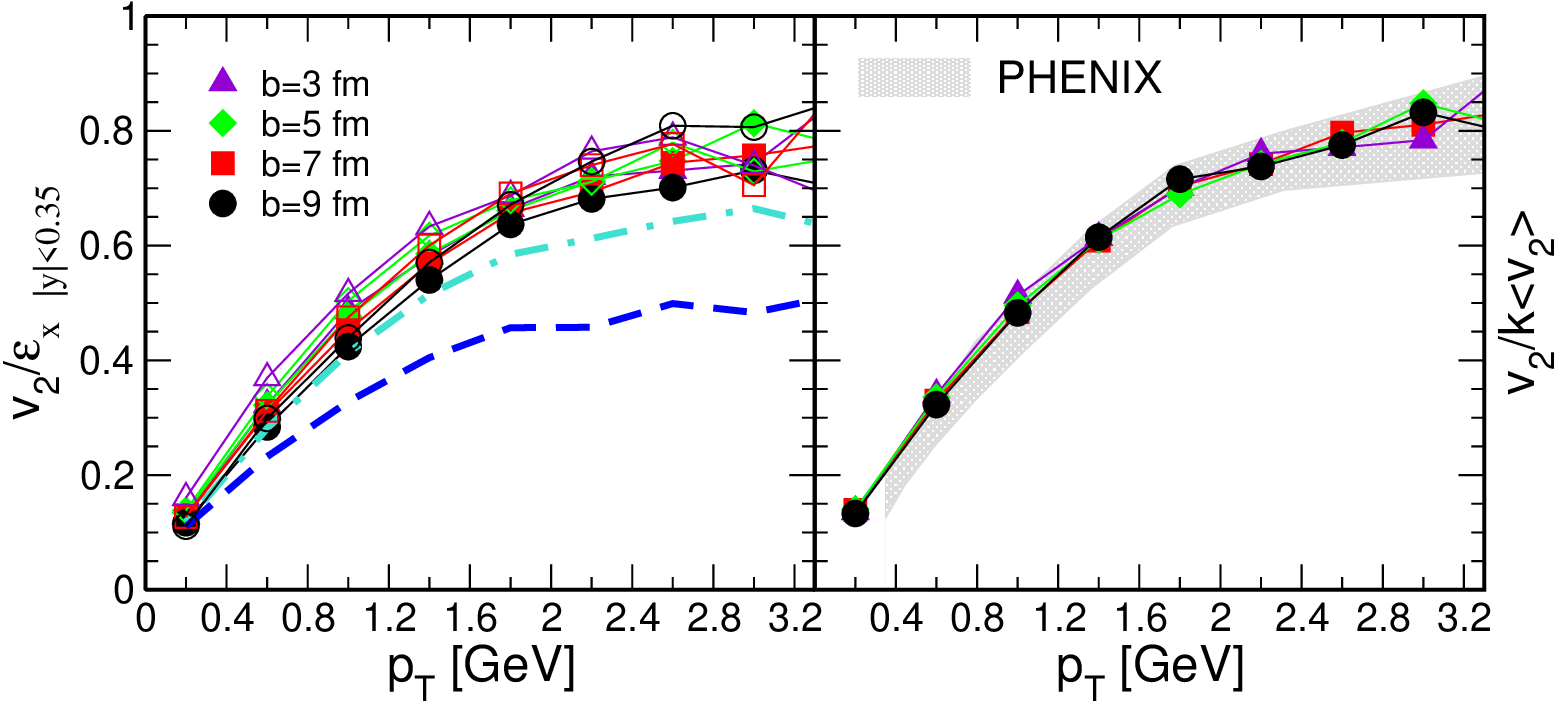}
\caption{$\frac{v_2}{\epsilon_x}$ (left panel)  and $\frac{v_2}{k\la v_2\ra}$ (right panel)
in the central rapidity region ($|y|<0.35$) for Au+Au (filled symbols) and Cu+Cu (open symbols) collisions at $\sqrt{s}=200 AGeV$. Different symbols refer to cascade simulations at various impact parameters for $\eta/s=1/4\pi$. In the left panel also results for 
Au+Au at b=7 fm with $\eta/s=1/2\pi$ (dot-dashed line) and $\eta/s=1/\pi$ (dashed line) are shown. The constant in the right panel is set to $k=3.1$ as in \cite{Adare:2006ti}.  }
\label{fig4}
\end{figure}

A first objective of our study is to investigate the scaling behavior of the elliptic flow with
the initial eccentricity and the system size to see if such a scaling typical of a hydrodynamical 
behavior \cite{Bhalerao:2005mm} persists also in a cascade approach. The interest for such a behavior is triggered by the
recent observation by the PHENIX Collaboration \cite{Adare:2006ti} of a scaling of $v_2(p_T)/\la v_2 \ra$ up to $p_T \sim 3$ GeV, a region usually considered
out of the range where hydrodynamics should work. 
In order to allow for a comparison with the results from ideal hydrodynamics of Ref.\cite{Bhalerao:2005mm} we have
followed the evolution of the system considering a finite constant viscosity, as in hydrodynamical
studies the zero mean free path condition is implicit for the entire evolution of the system.
The hadronic re-scattering together with the formation and decay of the resonances are neglected.
This is justified by the fact that the bulk of $v_2(p_T)$ develops in the early
stage of the reaction, i.e. well before hadronization sets in, as found by several theoretical approaches 
\cite{Zhang:1999rs,Lin:2001zk,Kolb:2003dz}
and more recently confirmed experimentally \cite{Afanasiev:2007tv,Abelev:2007rw}.
In Fig. \ref{fig4} the ${v_2}/{\epsilon_x}$ (left panel)  and ${v_2}/{k \la v_2\ra}$ (right panel)
in the central rapidity region ($|y|<0.35$) are shown
as a function of transverse momentum $p_T$ for different impact parameters and the two systems 
Au+Au (filled symbols) and Cu+Cu (open symbols) at 200 AGeV with our cascade approach when shear viscosity is kept constant at $1/4\pi$. The dot-dashed and dashed lines are the results for 
Au+Au at b=7 fm with $\eta/s=1/2\pi$ and $\eta/s=1/\pi$ respectively.
The value of the constant in the right panel is set to $k=3.1$, as in \cite{Adare:2006ti}, where $k\la v_2\ra$ is assumed to be equivalent to the initial eccentricity $\epsilon_x$.

A first important result is the clear observation of the
scaling both as a function of centrality and system size either for the 
$v_2(p_T)/\epsilon_x$ and the ${v_2(p_T)}/{\la v_2\ra}$. 
In fact, since the parton cross section is re-normalized in order to keep a small constant
shear viscosity, dynamical effects related to the different density and temperature conditions 
that are reached at the different impact parameters, are damped.
This indicates that the scaling $v_2(p_T)/\la v_2\ra$, which
is advocated as a signature of the hydrodynamical behavior \cite{Adare:2006ti} is a more general property that holds also
at finite mean free path or shear viscosity at least for values close to the lower bound. 
Moreover the scaling is shown to persist also at higher $p_T$ ($\sim$ 3 GeV) where
not only the scaling but also the saturation shape is correctly reproduced
by the parton cascade approach. 
Recently $v_2(p_T)$ has been investigated also with viscous hydrodynamics \cite{Romatschke:2007mq,Song:2007fn}. 
It is interesting that a similar (but weaker) $p_T$  dependence is found  with a quantitative agreement with minimum bias
data for $\eta/s \sim 0.1$. This is in general agreement with our calculations, performed at various impact parameters, 
as we can see in Fig. \ref{fig4} (right) comparing our results (symbols) with the shaded area (PHENIX data).
We however are not aware of an explicit investigation of $v_2(p_T)/\epsilon_x$ and $v_2(p_T)/\la v_2 \ra$ scaling within hydrodynamics.
Our simulations show a good sensitivity to the shear viscosity especially at intermediate $p_T$
where $v_2(p_T)/\epsilon_x$ drops of about $40\%$ when increasing $\eta/s$ by a factor 4 above the lower bound.
While the ${v_2}/{\la v_2\ra}$ scaling was initially considered to stand for the ${v_2}/{\epsilon_x}$ scaling and it is quoted as a further validation of ideal hydrodynamics \cite{Adare:2006ti}, 
latest results from STAR \cite{:2008ed} show that the $v_2(p_T)$ scaled by the participant eccentricity $\epsilon_x$ is not independent on centrality. In fact the build up of a stronger collective motion in more central Au+Au collisions is observed. On the other hand, a good scaling with centrality is instead observed for
${v_2}/{\la v_2\ra}$ ratio. This feature will be further clarified in the following.

\textbf{Effect of QGP freeze-out} - 
We investigate the effect of a freeze-out condition on the elliptic flow. Freeze-out conditions are justified by the
fact that, at a critical value for energy density, hadronization sets in and parton dynamics is no longer 
acting. To take into account such an effect we stop
the interactions among partons as the local energy density drops below 0.7 GeV/fm$^3$, an intermediate value in the range corresponding to a mixed quark-hadron phase \cite{Kolb:2003dz}. 
Previous calculations were performed with a freeze-out condition at $\epsilon = 0.2$ GeV/fm$^3$
which corresponds to the end of a mixed phase or roughly to an hadronic-thermal freeze-out.
We have checked that
a freeze-out at $0.2$ GeV/fm$^3$ is practically identical to consider no energy density freeze-out at all.
This is in agreement with the observation that
both theoretically and experimentally the elliptic flow does not develop significantly during the hadronic stage
\cite{Kolb:2003dz,Abelev:2007rw,Zhang:1999rs,Lin:2001zk,Afanasiev:2007tv} ,
hence we will refer to such a calculation as the one without freeze-out. 
\begin{figure}[ht]
\includegraphics[height=1.7in,width=2.6in]{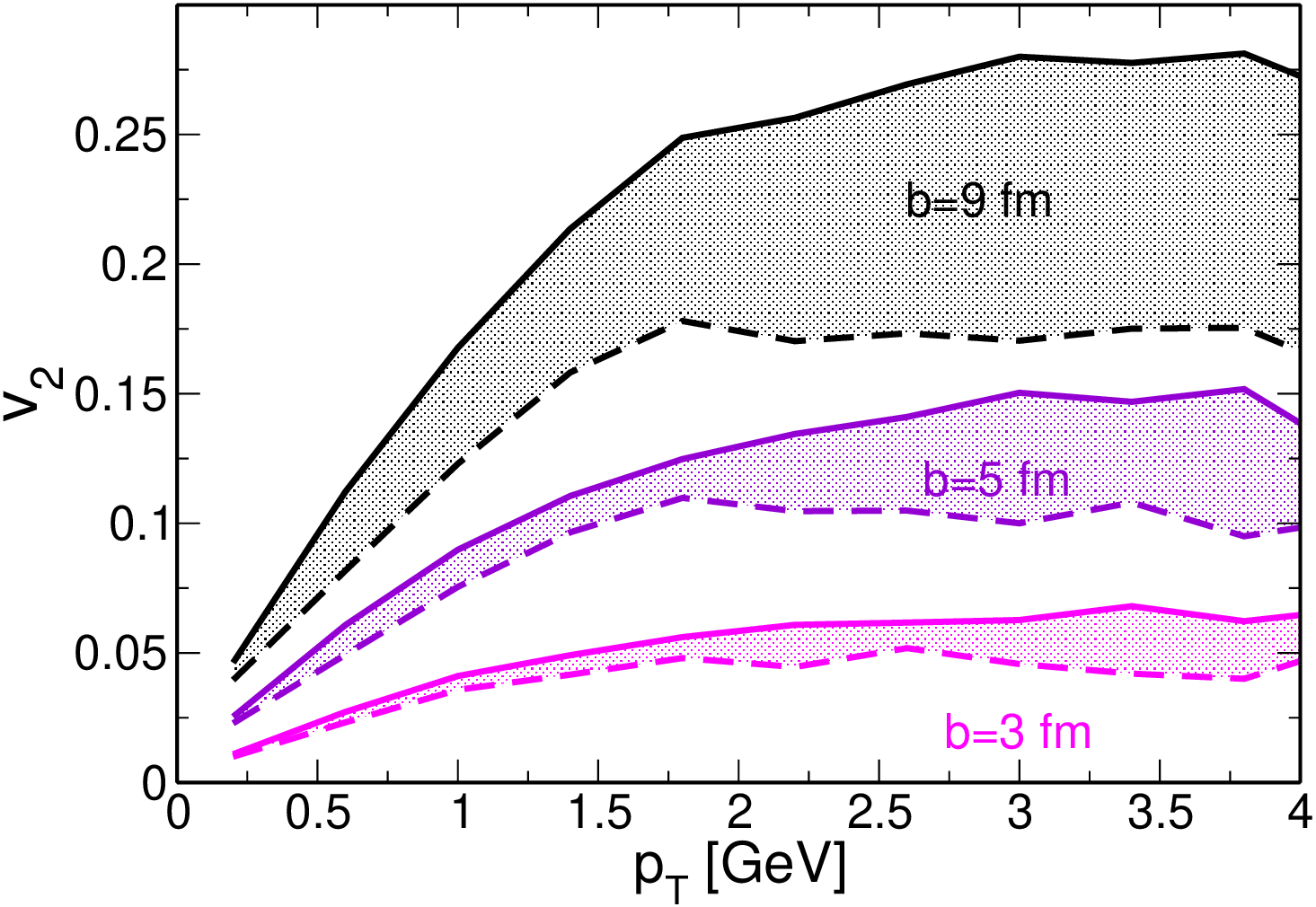}
\caption{Differential elliptic flow for Au+Au at different impact parameters with (dashed curves) and without (solid curves) a freeze-out condition ($\epsilon_{crit}=0.7$ GeV fm$^{-3}$).}
\label{fig5}
\end{figure}

\begin{figure}[ht]
\includegraphics[height=1.7in,width=3.3in]{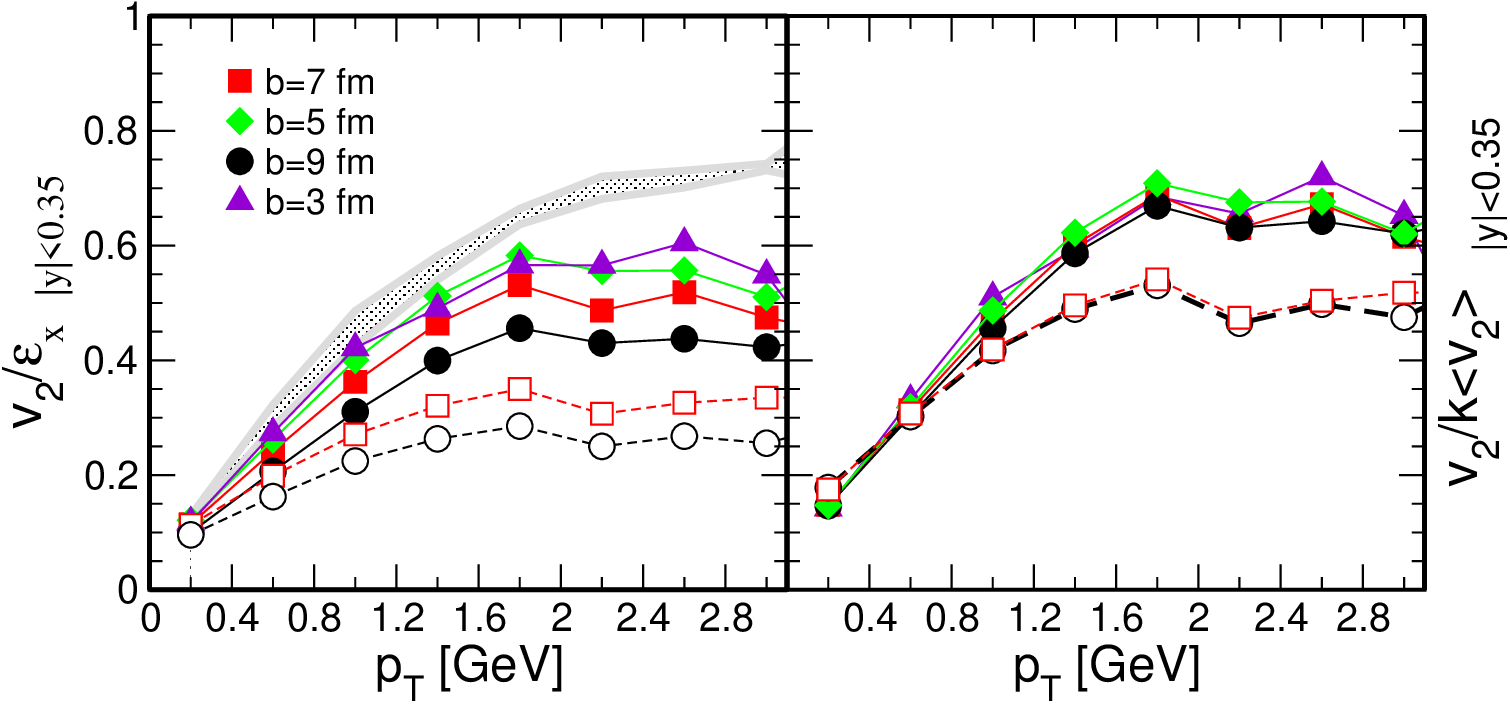}
\caption{Same as Fig.\ref{fig4}, filled symbols refer to calculations at $\eta/s=1/4\pi$; open symbols are for b=7 fm (squares) and b=9 fm (circles) calculations at $\eta/s=1/\pi$. The grey band in the left panel refers to the results of simulations without freeze out (see left panel in Fig. \ref{fig4}).  }
\label{fig6}
\end{figure}

When the freeze-out condition is implemented a sizeable reduction for the elliptic flow is observed (see Fig. \ref{fig5}), especially for the most peripheral collisions and at larger $p_T$. 
Correspondingly our results show that the scaling of elliptic flow with the initial spatial eccentricity is broken 
(see filled symbols in left panel of Fig. \ref{fig6}). In particular $v_2/\epsilon_x$ varies of nearly $40-50\%$ from b=3 fm to b=9 fm in the intermediate $p_T$ region ($<$3 GeV). The amount of such a spreading is consistent with the data reported by \cite{:2008ed} for the centrality selections $0-10\%$ and $10-40\%$, with central collisions exhibiting a bigger elliptic flow to eccentricity ratio than the peripheral ones. On the other hand, the scaling of $v_2/\la v_2\ra$ with the impact parameter is still observed (see right panel in Fig. \ref{fig6}). 
We are therefore driven to the conclusion that the breaking of the 
${v_2(p_T)}/{\epsilon}$ scaling, as observed in Fig. \ref{fig6},
traces back to the freeze out physics, which deserves a deeper investigation.

\begin{figure}[ht]
\includegraphics[height=1.7in,width=2.5in]{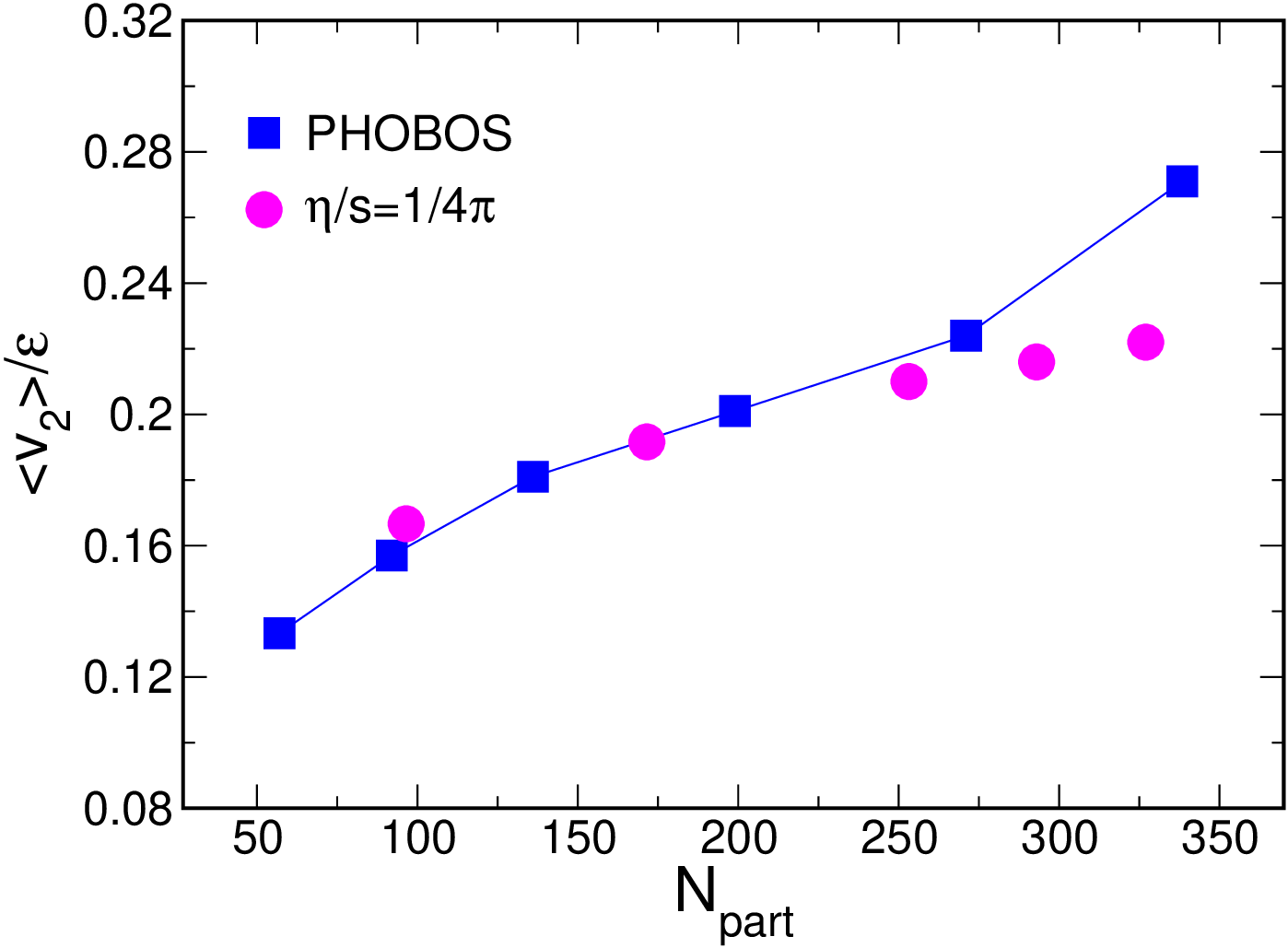}
\caption{$\frac{\la v_2\ra}{\epsilon}$ as a function of participant number for Au+Au collisions at $\sqrt{s}=200$ GeV in the central rapidity region $|y|\leq 1$. Cascade results for $\eta/s = 1/4\pi$ are divided by 1.2. Squares are the corresponding data from \cite{Alver:2006wh}.}
\label{fig1}
\end{figure}

We are mainly focused on the $p_T$ dependence but it is of course interesting
to look at the behavior of the averaged $\la v_2 \ra$.
We however note that a comparison of the averaged $v_2$ in our parton cascade with experimental data
should face two main limitations. One is the lack of finite quark mass
effect that are known to reduce the value of $v_2$ up to a $p_T$ of the order of the mass, an effect known in a
hydrodynamical picture as mass ordering. The other limitation is due to the absence of resonance formation
and decay which are known to affect the elliptic flow especially for pions. 
Both effects are relevant at $p_T < 1$ GeV and therefore should be taken into
account to estimate the absolute value of $\la v_2\ra$, as appropriately
done in hydrodynamical models. Nevertheless we are mainly interested in showing that
a cascade approach is able to reproduce the observed trend of  $\la v_2\ra/\epsilon_x$ with the number of participant $N_{part}$.
Within the ideal hydrodynamics picture, the $p_T$ averaged elliptic flow is approximately
proportional to the initial spatial eccentricity $\epsilon_x$, leading to a 
centrality independent value of the ${\la v_2\ra}/{\epsilon_x}$ ratio. However recent measurements
performed by STAR and PHOBOS \cite{Alver:2006wh,:2008ed} have pointed out significant deviations
from the scaling. 
The trend experimentally observed (squares in Fig.\ref{fig1}) is recovered in our cascade approach at finite viscosity
together with the scaling of $v_2(p_T)/\la v_2\ra$. 
To compare with the experimental data we divide our results by a factor 1.2, which
corresponds to fit the experimental value at $N_{part}\simeq170$, see Fig.\ref{fig1}. 
We mention that a breaking of the scaling for the average $\la v_2 \ra$ is seen also without freeze-out
condition, even if the effect of freeze-out reduces the absolute value and enhances the breaking.

\begin{figure}[ht]
\includegraphics[height=1.7in,width=2.5in]{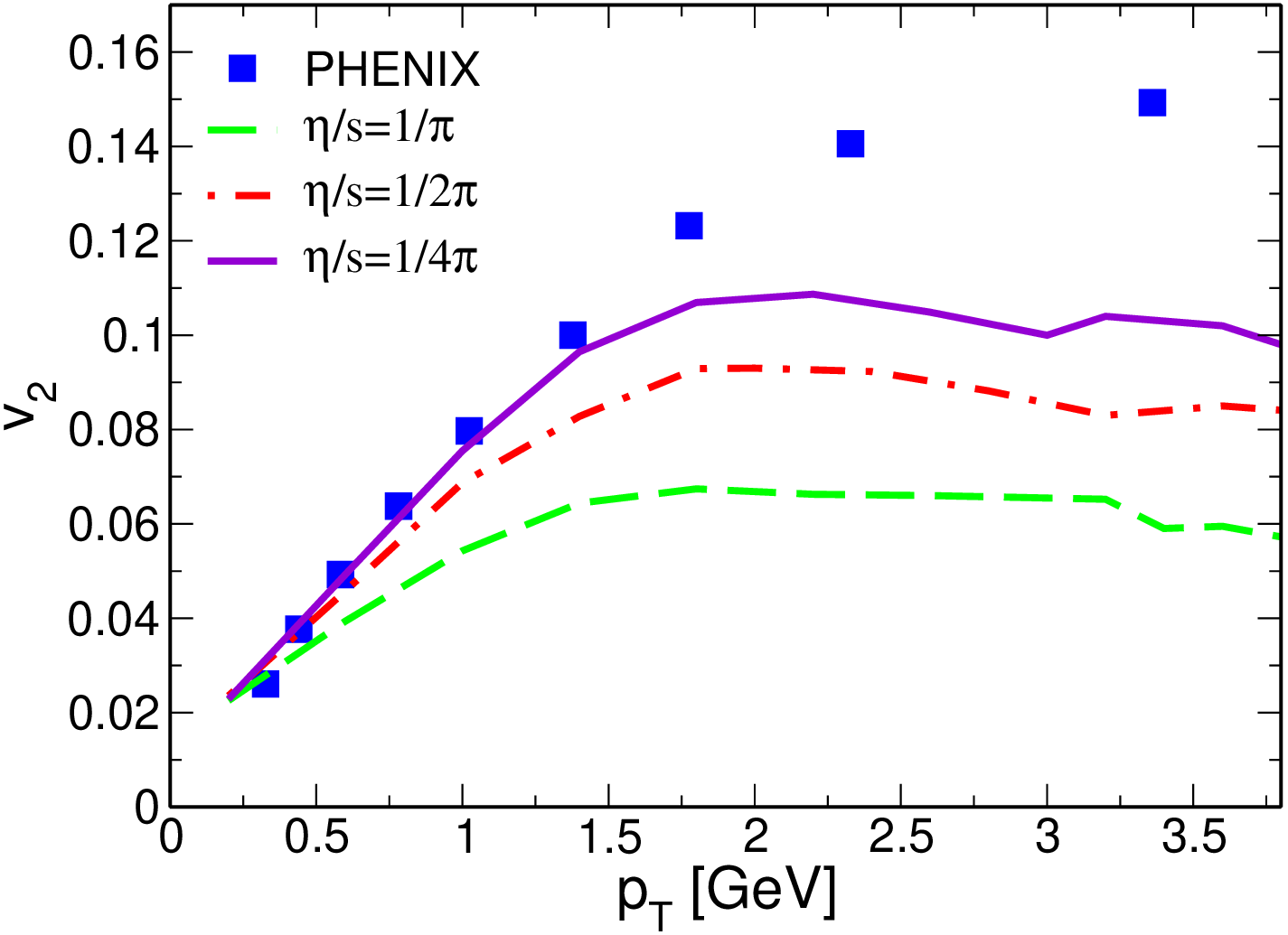}
\caption{Differential elliptic flow for Au+Au collisions at b=5 fm, $|y|\leq 0.35$ and $\eta/s=1/4\pi$ (solid curve), $\eta/s=1/2\pi$ (dot-dashed curve) and $\eta/s=1/\pi$ (dashed curve). Results from cascade are compared with data from \cite{Adare:2006ti} (squares). }
\label{fig7}
\end{figure}

As a last point we discuss how information on the viscosity of the RHIC plasma can be inferred from the $v_2(p_T)$ absolute value. 
One has to consider that at intermediate $p_T$ there are several evidences
for hadronization via coalescence and it has been shown that due to a coalescence mechanism the parton $v_2$ translates into a nearly doubled hadron $v_2$ \cite{Greco:2007nu,Annual},
therefore a definite evaluation of $\eta/s$ from $v_2(p_T)$ data needs to include the coalescence plus fragmentation
mechanism that should account for the baryon-meson quark number scaling. We refrain from using simple naive
coalescence formula \cite{Kolb:2004gi} here, considering that it has been shown that space-momentum correlation 
and the freeze-out hypersurface can significantly affect the relation between quark and hadron $v_2$  \cite{Pratt:2004zq,Molnar:2004rr,Greco:2005jk}.
It is therefore necessary a further development of the parton cascade approach that includes self-consistently
the coalescence and fragmentation process.
Nonetheless from Fig.\ref{fig7} we notice that for $\eta/s=\pi^{-1}$ the parton elliptic flow with
a quite small slope at low $p_T$ saturates at about $6\%$. Even assuming a coalescence mechanism in the hadronization phase, this value appears to be too low to reproduce the baryon and meson $v_2$. This provides anyway an indication that a shear viscosity as high as 4 times the minimum value should be ruled out for the RHIC fluid and the viscosity is therefore quite smaller than pQCD estimates \cite{Arnold:2003zc}.
On the other hand the results with both $\eta/s=1/4\pi$ and $\eta/s=1/2\pi$ could be quite close to the experimental data within a coalescence picture. Such a range of values, to be narrowed in the next future, is sligthly larger than the first estimates with viscous hydrodynamics \cite{Romatschke:2007mq,Song:2007fn}, slightly below to the one based on
Knudsen number analysis \cite{Drescher:2007cd} and contains the best present evaluation in lQCD \cite{Meyer:2007ic}.

\textbf{Summary and Conclusions} - 
We have investigated the dependence on the shear viscosity of the elliptic flow $v_2(p_T)$ and its
scaling properties. As a first result we find that the approximate scaling of $v_2(p_T)/\epsilon_x$
advocated as a signature of the perfect hydrodynamical behavior \cite{Adare:2006ti} can still hold also at
finite viscosity and in a parton cascade approach. However such a scaling versus centrality
and system size is present only if one 
makes the fireball evolve down to energy density $\epsilon \sim 0.2$ GeV/fm$^3$ corresponding 
typically to the end of a mixed phase. If a freeze-out condition for the partonic dynamics is put at
$\epsilon \sim 0.7$ GeV/fm$^3$ then a sizeable breaking of the $v_2(p_T)/\epsilon_x$ scaling is seen while
$v_2(p_T)/\la v_2\ra$ still scales. This is in qualitative agreement with
the recent experimental data from STAR \cite{:2008ed} indicating that freeze-out of QGP dynamics 
with the consequent change of $\eta/s$ should be more thoroughly investigated.
As a final remark we notice that without any freeze-out condition the $v_2(p_T)$ at parton level
would be close to the data for $\eta/s =1/4\pi$, see Fig.\ref{fig1}. On the other hand we consider
such an agreement misleading because it would not be compatible with the enhancement
of $v_2$ due to coalescence and the observation of quark number
scaling \cite{Annual}. Instead the freeze-out condition seems to pave the way for a consistency among the different
available observables on elliptic flow: the breaking of $v_2(p_T)/\epsilon_x$, the persistence of
$v_2(p_T)/\la v_2\ra$ scaling and the presence of a coalescence plus fragmentation hadronization
mechanism acting at intermediate $p_T$.
We therefore conclude that a safe evaluation of shear viscosity from the available data on $v_2(p_T)$ 
necessitates a cascade approach that includes self-consistently hadronization by coalescence
and fragmentation.
Finally we mention that such an investigation can be strengthened by a study of the fourth harmonic
in the azimuthal anisotropy, i.e. the $v_4=\la cos (4 \phi) \ra$. A first analysis shows a stronger
sensitivity to $\eta/s$ and especially a more critical dependence on the freeze-out dynamics 
respect to $v_2$\cite{tokai}.

\end{document}